# Absolute Value Constraint: The Reason for Invalid Performance Evaluation Results of Neural Network Models for Stock Price Prediction

Yi Wei
Department of Computer Science and Engineering
University at Buffalo
Buffalo, NY, USA 14260
yiwei@buffalo.edu

**Abstract**

Neural networks for stock price prediction(NNSPP) have been popular for decades. However, most of its study results remain in the research paper and cannot truly play a role in the securities market. One of the main reasons leading to this situation is that the prediction error(PE) based evaluation results have statistical flaws. Its prediction results cannot represent the most critical financial direction attributes. So it cannot provide investors with convincing, interpretable, and consistent model performance evaluation results for practical applications in the securities market. To illustrate, we have used data selected from 20 stock datasets over six years from the Shanghai and Shenzhen stock market in China, and 20 stock datasets from NASDAQ and NYSE in the USA. We implement six shallow and deep neural networks to predict stock prices and use four prediction error measures for evaluation. The results show that the prediction error value only partially reflects the model accuracy of the stock price prediction, and cannot reflect the change in the direction of the model predicted stock price. This characteristic determines that PE is not suitable as an evaluation indicator of NNSPP. Otherwise, it will bring huge potential risks to investors. Therefore, this paper establishes an experiment platform to confirm that the PE method is not suitable for the NNSPP evaluation, and provides a theoretical basis for the necessity of creating a new NNSPP evaluation method in the future.

## 1. INTRODUCTION

### 1.1 Performance Evaluation of NNSPP

The existing prediction task of the stock price using neural networks is implemented by training the model parameters to obtain the minimum training error, and evaluated on the test dataset by comparing their prediction errors(test error). It is widely used as the criterion for model comparison and becomes an important indicator for the performance of stock price prediction[1][2]. It is important to prove whether the result of a stock price prediction comes from the model's contribution, rather than the interference due to factors such as the distinctions between stock price datasets. The implementation of stock price prediction is directly market-oriented. The quality of its measurement is based on whether the prediction results can effectively help investors profit or stop loss, rather than the high and low value of the model's prediction error. This issue was noticed by researchers as early as 1988 [3].

Many neural networks for stock price prediction miserably failed when applied in actual trading [4]. This shows that the application effect of artificial intelligence tools is not as simple as assessing the superiority based on the prediction error alone [5][6][7]. Investors, in particular, need more than the performance from a particular prediction algorithm. Besides performance, there is one other basic indicator for the practicability of neural networks that is very important: the completeness of functionality. Currently, the evaluation of stock price prediction models mainly focuses on their performance, and there is no direct test of their functionality. Therefore, the "use value" of neural networks is doubtful for stock price prediction. To a stock investor, what is needed is a stock price prediction model that can help them outperform the market index and outperform a target stock, rather than several machine learning products with low prediction errors but little directional prediction ability. It is important for researchers to ensure that NNSPP is aligned with the actual trading needs.

The performance of NNSPP is evaluated based on the prediction error metric [8], and this evaluation method has statistical flaws for the securities market. As we all know, the PE metric takes the absolute value of the prediction error, while the stock price movements are directional. In short, they are either rising or falling, which is the most basic fractal feature of financial time series [9][10]. Many researchers have apparently neglected the effectiveness of PE in evaluating the performance of regression neural networks. Strictly speaking, the PE method does not have the basic function of evaluating financial time series because the same level prediction error cannot reflect the general movement direction of the financial time series being evaluated. Therefore, in practical applications, the technique will mislead investors and potentially cause major economic losses.

### 1.2 Interpretability of Neural Network

In the field of machine learning, the interpretability of models has significant theoretical and practical value. Machine learning models with high interpretability are more trustworthy and more likely to be accepted by users. The development of machine learning algorithms has accompanied the study of interpretability. Early machine learning algorithms were relatively simple, and as time went on, researchers continuously proposed various advanced machine learning models. However, as the complexity and accuracy of the model both increase, its interpretability dramatically decreases. Although deep neural networks have demonstrated powerful performance in applications such as image, speech, and video [11][12], the models it trains are usually complex and difficult to understand. Consequently, their defects in terms of interpretability have been criticized by researchers and users [13][14].

The application fields of interpretability research are very wide. Thus, the interpretability research of many machine learning algorithms combine characteristics of their respective applications, for example, in the field of image recognition [15], text processing [16][17]. Besides, there are many interpretability related research achievements in the fields of computer-aided design [18], education [19], neuroimaging [20], psychology [21], pharmacy [22], art [23], etc.

As we all know, whether a neural network is practical or not, there are two basic indicators that are very important: one is the completeness of functionality; the other is the criteria of prediction performance. The evaluation of NNSPP mainly focuses on their performance at present, and there is no in-depth study on the applicability of their evaluation standards in the securities market. This situation makes the practical value of NNSPP very uncertain. For stock investors, what is needed is that NNSPP's prediction results can help them make correct trading decisions. However, the current evaluation results using the PE method in NNSPP cannot meet the above-mentioned interpretability requirements.

Because of the strong demand for the model interpretability in practical applications, researchers pay increasingly more attention to it. By searching articles about model interpretability in the web of science databases, it can be found that the total number of publications in this field and the number of paper citations have been steadily increasing in recent years. Besides, the International Conference on Machine Learning has been hosted a Workshop on Human Interpretability in Machine Learning each year since 2016. On the other hand, the demand for model interpretability has risen to a national legal and strategic level. At present, interpretability is becoming a hurdle for AI. On December 18, 2018, the European Union issued the "Artificial Intelligence Ethics Guidelines" and "Ethics Guidelines for Trustworthy AI," which clearly stated that the direction of AI development should be trustworthy, safe, transparent, and interpretable.

## 2. Background

### 2.1 The Problem Discovery of PE's Functional Drawback for NNSPP Performance Evaluation

The classical NNSPP performance evaluation is determined based on the high and low value of the prediction error [7][24][26], but for the securities market, this evaluation method has statistical flaws. As we all know, the PE metric uses the absolute value of the prediction error. The trend of stock prices is directional-whether they rise or fall is the most basic fractal feature of financial time series. However, many researchers unconsciously neglect the securities trading effectiveness of the PE method in the performance evaluation results of neural networks. Thus, in practical applications, the value of prediction error will mislead investors and inevitably cause significant economic losses. The preliminary work of this paper shows that PE does not have the basic function of evaluating NNSPP, because the same level of forecast error cannot reflect the direction of movement of the evaluated financial time series [25]. Moreover, this feature is crucial for the performance characterization of NNSPP. The users of the PE method have not noticed whether the value of the prediction error can effectively help model users. In other words, when the existing PE method is used by people in the performance evaluation of NNSPP, the value of prediction error is just an evaluation index deceiving themselves as well as others. Obviously, one of the reasons for this situation is that the model evaluator lacks professional knowledge of securities trading. The second reason is that the developed model cannot be used in actual stock trading. Therefore, the existing PE method in evaluating the functional defects of NNSPP has not been recognized for a long time. Objectively speaking, the PE method only partially evaluates the performance of NNSPP, not the overall performance. White pointed out as early as 1988 that the performance of NNSPP cannot be well evaluated by the value of prediction error alone [3]. Besides, some researchers also proposed the importance of direction notation in studying financial time series [27]. Unfortunately, many NNSPP developers did not pay attention to these useful suggestions. In the preliminary work of this research, Dr. S. L. Wei, an investor with more than 20 years of experience in securities trading, when assisting this research in evaluating the trading performance of NNSPP, once pointed out: "The PE method used to evaluate the performance of NNSPP, has statistical flaws for securities trading, and its evaluation results are very misleading." At present, Wei and Chaudhary have preliminarily confirmed the correctness of this argument [25]. In summary, from the perspective of the securities profession, the PE method is not suitable for the performance evaluation of NNSPP. Its only benefits are to make those research papers appear more rigorous and convincing, nothing more. The raising of this question makes the applicability of the existing PE method in NNSPP performance evaluation facing serious challenges!

### 2.2 The Potential Dangers of Existing NNSPP Performance Evaluation Methods to Securities Investors

Swing Trading of stocks refers to an investment method in which people who invest in stocks will sell high and buy low. The visual positive and negative price in the graph corresponding to the rising and falling of stocks is usually called the stock price direction. Elliott Wave Theory believes that the rising and falling of stock prices are the primary forms of stock waves. Effectively grasping the rising and falling stock trends is the fundamental guarantee for winning in the stock market. This universal core viewpoint has long been regarded as a classic by stock market professionals [28]. The stock price sequence's fundamental element is rising and falling, and the current NNSPP performance evaluation does not notice this feature. The existing PE method emphasizes the accuracy of predicting stock prices while ignoring the critical issue of characterizing the direction of the stock price sequence [25]. As a result, its evaluation results are not interpretable and cannot be used to guide securities trading.

The application effect of AI tools such as neural networks for stock price prediction is not as simple as assessing its pros and cons based on the high and low value of prediction error [5][6][7]. Moreover, for investors, the problems faced by this NNSPP are not just the question of its performance, but also the question of whether its evaluation method is reliable and whether the evaluation result can guide securities trading. About the problems mentioned above, it is not that it has not been taken seriously by some professionals [3], but it has not attracted enough attention.

From the perspective of securities investors, the accuracy of the NNSPP prediction error is the second most important. In other words, without the control variable of ups and downs, the traditional prediction error can only be regarded as a meaningless ornament. That is to say, if the NNSPP performance evaluation method cannot meet the professional needs of the NNSPP performance evaluation, which is the first property of fluctuations and the second is accurate, it is not very sensible. Suppose we switch the application scenario to the securities market. In that case, we will also come to the same conclusion: the prediction error value without directional attributes will not be of any practical value to securities investors except for misleading them.

The lack of no-treatment-control to test the effectiveness of NNSPP's evaluation results is the objective reason for the failure to find the above technical defects in the PE method for a long time. In actual securities trading, if the NNSPP, whose evaluation result cannot reflect its true performance, is applied, investors will lose confidence in the predicted results of the model. Because only considering the level of the prediction error of the model without considering the evaluation result of the actual rising and falling of the predicted stocks, it will inevitably mislead stock traders, which will expose them to substantial potential financial risks.

## 3. METHODOLOGY

### 3.1 Financial Terminology

This definition of financial terminology used in this paper are as follows:

**Absolute Value Constraint:** It is the abbreviation for the absolute value constraint of the stock rising and falling. The performance evaluation result of the existing NNSPP is based on the prediction error's value to determine the model's quality. Because the PE output data is an absolute value, it cannot represent the stock price sequence's original rising and falling attributes. This phenomenon is called the absolute value constraint of the stock rising and falling trend. This feature determines that the existing NNSPP's PE evaluation method can only be used for the AI model research, the non-financial performance evaluation.

**Control Variable:** Also known as additional related variables, refer to the potential factors or conditions that affect the experiment's changes and results in addition to the experimental variables.

**Interpretability:** There is currently no mathematical definition of interpretability. Interpretability in artificial intelligence means that we have enough understandable information to solve a problem and provide a decision basis for each predicted result [29].

**Invalidity of Securities Trading:** It means that the methods or data used have no meaning for securities trading and cannot be used in securities trading. For example, the existing PE method's evaluation result of NNSPP cannot reflect the predicted stock price sequence's ups and downs. It cannot be used as an operating basis for stock trading. Therefore, securities professionals believe that the existing PE method's evaluation result of NNSPP in securities trading is helpless.

**NNSPP**: Neural Network for Stock Price Prediction

**No-treatment-control**: It refers to the control group selected in the experiment without any treatment. The significance of setting a no-treatment-control group is to eliminate the interference of other factors on the test results. Ensure the credibility of test results and improve the reliability of test results[30].

**PE**: Prediction Error

**Rate of Stock Return**: The ratio of the total amount of net gain on investment to the investment's initial cost.

**Stock Return**: It refers to the capital gains, which are the appreciation gains of stocks, such as the stock price difference after buying at a lower price and selling at a higher price.

**Stock Return Direction**: It refers to, within a certain period of time, the positive and negative return status corresponding to the rising and falling of the stock. In actual securities trading, the stock return direction is what investors usually call the rising and falling direction of stocks.

### 3.2 Stock Data

The training dataset used in this paper is 20 stocks in the Shanghai and Shenzhen stock market from China, and 20 stocks in NASDAQ and S&P500 from the US stock market. The data are collected from the Yahoo Finance website (http://finance.yahoo.com/). The data range is from 2011-01-01 to 2017-06-02. The dataset has a total of 62,040 trading days. During training, stock data from 2011-01-01 to 2016-12-31 is used as the training dataset, and stock data from 2017-01-10 to 2017-06-02 as the testing dataset. The time window based historical price is used as the input data, and the actual price of each corresponding window is used as the target data. Besides, to ensure the stock dataset mining quality, the stock dataset must also be verified by fractal trend before use, which is to calculate their Hurst exponent. The integrity verification and normalization of stock price data are also necessary to implement.

### 3.3 Neural Networks

Six types of the neural network including Multilayer Perceptron (MLP), Recurrent Neural Network (RNN), Long Short Term Memory network(LSTM), Gated Recurrent Unit network(GRU), Bidirectional Recurrent Neural Network(BiRNN), and Bidirectional Long Short Term Memory network(BiLSTM) are employed to conduct predictions.

## 3.4 Parameter Design

### 3.4.1 Parameter Optimization Design

Stock price prediction task will involve multi-factor experiments. In order to avoid interference due to subjective design factors and blindly increasing the number of trials, it is necessary to optimize the design of experimental parameters and meanwhile to save manpower and time. This can reduce the number of trial times and ensure that the test results are accurate and reliable to efficiently obtain scientific conclusions from overall tests.

At present, the widely used experimental optimization design method is orthogonal test design. It uses statistics, probability theory, and practical needs as the starting point. It uses orthogonal tables to formulate the test plan, and then uses the analytical method to analyze the results to obtain optimal test conditions.

Orthogonal experiment design is a statistical experiment design method. It can select some representative factors from the comprehensive experiment for model training. It is a scientific design optimization method that can quickly deal with complex problems. This method achieves orthogonality, balanced dispersion, and neat comparability of experimental factors through the design of orthogonal tables. It has become an effective tool for scientific experiments [31][32][33][34].

### 3.4.2 Orthogonal Experiment Design

In this study, the sequence of the stock price is created by the sliding window approach, and the short-term univariate prediction of the stock price is investigated via neural networks. The parameters are selected by orthogonal experiment design optimization for different factors in the training process, including the activation function, the number of hidden nodes, and the observation window size and prediction window size as part of the training data pre-processing procedure, etc. Through the design of $L_{16}(4^5)$, the orthogonal array testing is conducted to reduce test cases from 1024 to 16 in order to scientifically and effectively realize the multi-level optimization process of multiple factors during training.

Table 3.1 The factors and levels for orthogonal array testing

| Factor | Level | | | |
|---|---|---|---|---|
| Window Length (**L**) | 5 | 10 | 15 | 20 |
| Window Hop Count (**H**) | 0 | 1 | 2 | 3 |
| Number of Hidden Node (**N**) | 5 | 10 | 15 | 20 |
| Epoch (**E**) | 10 | 50 | 200 | 1000 |
| Activation Function (**F**) | Linear | Sigmoid | Tanh | ReLU |

In order to fully demonstrate the prediction effect, five commonly used parameters in this study are selected for orthogonal array testing. Four levels of each factor are used to explicitly show their influence on the prediction. Through a table look-up, the experimental scheme $L_{16}(4^5)$ is selected with four levels of five factors: the 16 different combinations will be performed for each stock (Table 3.1).

## 3.5 Prediction Error Evaluation

There are many methods to check the prediction accuracy of the regression neural network. In this study, four methods of mean absolute error (MAE), mean square error (MSE), root mean square error (RMSE), and $R^2$ are used to verify the accuracy of stock price predictions of different neural networks.

### 3.5.1 Mean Squared Error (MSE)

$$MSE = \frac{1}{N}\sum_{i=1}^{n}(y_i - \hat{y}_i)^2 \qquad (3.1)$$

Where $y_i$ is the actual output, and $\hat{y}_i$ is the model's prediction. MSE measures the mean square error of the prediction. For each time point, it calculates the squared difference between the predicted value and the target value, and then takes the average. The higher the value, the worse the model. But MSE has the disadvantage of greatly weighting outliers [35]. If a very bad prediction is made, the square will make the error more serious and may bias the indicator to overestimate the defects of the model.

### 3.5.2 Root Mean Squared Error (RMSE)

RMSE is just the square root of MSE. Introducing the square root can make the error range the same as the target range. Therefore, even in terms of model ranking, RMSE and MSE are indeed very similar, but for gradient-based methods, they cannot be interchanged with each other. Therefore, RMSE will also be used in this study.

$$RMSE = \sqrt{MSE} = \sqrt{\frac{1}{N}\sum_{i=1}^{n}(y_i - \hat{y}_i)^2} \qquad (3.2)$$

In formula 3.10, N is the number of samples, $\hat{y}_i$ is the i-th predicted value, and $y_i$ is the *i*-th observed value.

RMSE is used to aggregate the size of prediction errors at different times into a single prediction capability metric. Like MSE, it is a scale-dependent prediction error measurement method [36], which is used to compare the prediction errors of a specific dataset rather than different datasets between models.

### 3.5.3 Mean Absolute Error (MAE)

MAE calculates the average value of the absolute difference between the target value and the predicted value. MAE measures all individual differences and is evenly weighted in the average. But RMSE is not. Mathematically, it is calculated using the following formula:

$$\text{MAE} = \frac{1}{N}\sum_{i=1}^{n}|y_i - \hat{y}_i| \qquad (3.3)$$

The significance of this indicator is that it can punish huge errors, and such errors will not be as serious as MSE. Therefore, it is less sensitive to the mean square error with outliers [37]. The mean absolute error (MAE) uses the same scale as the measured data. This is called a scale-related accuracy measure and, therefore, cannot be used to compare between series using different scales [38].

### 3.5.4 R-Squared (R²)

It is difficult to know whether a model is accurate by looking at the absolute value of MSE or RMSE. We may want to measure the degree to which the model is better than a constant benchmark. It provides a way to measure the degree to which the model replicates observations based on the proportion of the results interpreted by the model in the total change of results [39].

The determination coefficient or $R^2$ is another indicator that we can use to evaluate the model. It is closely related to MSE. It has the advantage of non-scale and does not matter whether the output value is very large or very small. $R^2$ is always between $-\infty$ and 1. between.

When $R^2$ is negative, it means that the model is worse than predicting the mean.

$$R^2 = 1 - \frac{MSE(model)}{MSE(baseline)} \quad (3.4)$$

The MSE of the model is computed as above, while the MSE of the baseline is defined as:

$$MSE(baseline) = \frac{1}{N}\sum_{i=1}^{n}(y_i - \bar{y})^2 \quad (3.5)$$

Where the $y$ with a bar is the mean of the observed $y_i$.

More specifically, this baseline MSE can be regarded as the MSE that the simplest model may obtain. The simplest model may always predict the average of all samples. A value close to one indicates that the model error is close to zero, and a value close to zero indicates that the model is very close to the baseline.

In investments, R-squared is usually interpreted as the percentage of changes in the prices of funds or securities that can be explained by changes in the benchmark index. This can be applied to the comparison of stocks with the S&P 500 index or any other relevant equity index.

### 3.6 Stock Return Calculation

Define a test dataset interval of length n, $S_n = \{p_{t-n}, p_{t-n-1}, \ldots p_{t-1}, p_t\}$

Range span return represents the overall gain of $S_n$

$$R_{S_n} = \frac{p_t - p_{t-n}}{p_{t-n}} \quad (3.6)$$

### 3.7 Correlation Analysis

Pearson correlation coefficient is an evaluation standard used to measure the degree of linear correlation of variables or vectors [40]. The Pearson correlation coefficient is always between $-1.0$ and $1.0$. If the value is greater than 0, it indicates a positive correlation, and less than 0 indicates a negative correlation. A large absolute value means a strong correlation between variables or vectors. Close to 0 means the correlation between variables is small, almost no correlation. Pearson's correlation coefficient is represented by $\rho_{x,y}$. The calculation method is to compare the covariance and standard deviation between two variables. The formula is as follows:

$$\rho_{x,y} = \frac{cov(X,Y)}{\sigma_x \sigma_y} = \frac{E((X-\mu_X)(Y-\mu_Y))}{\sigma_x \sigma_y} \quad (3.7)$$

Where x and y are two variables, and N is the total number of samples.

The definition of correlation rules to evaluate stock price prediction is as follows:

Table 3.2 Rule of Thumb for Interpreting the Correlation Coefficient [41]

| Size of Correlation $\rho_{x,y}$ | Interpretation |
|---|---|
| $0.9 < \|\rho_{x,y}\| \leq 1.0$ | Very high positive (negative) correlation |
| $0.7 < \|\rho_{x,y}\| \leq 0.9$ | High positive (negative) correlation |
| $0.5 < \|\rho_{x,y}\| \leq 0.7$ | Moderate positive (negative) correlation |
| $0.3 < \|\rho_{x,y}\| \leq 0.5$ | Low positive (negative) correlation |
| $0.0 < \|r\rho_{x,y}\| \leq 0.3$ | Negligible Correlation |

## 4. RESULT ANALYSIS

### 4.1 Performance Evaluation of the NNSPP

#### 4.1.1 Model prediction error of different stock datasets

Traditionally, prediction error has been an essential measure of the performance evaluation of NNSPP. In this study, MAE was used to examine the prediction error of 40 stock datasets. Taking the BiLSTM network as an example, comparing the prediction error among 40 datasets, 300027 has the smallest prediction error of 0.0133 in the Chinese stock dataset, and AMZN has the smallest prediction error of 0.015 in the US stock dataset (Table 4.1).

Table 4.1 Performance evaluation of NNSPP using MAE prediction

| Stock Symbol | MAE | Return Rate(%) | Stock Symbol | MAE | Return Rate(%) |
|---|---|---|---|---|---|
| 000002 | 0.0290 | 1.80 | AAPL | 0.0240 | 30.51 |
| 000014 | 0.0387 | -17.74 | BA | 0.0188 | 19.59 |
| 000538 | 0.0362 | 21.06 | BAC | 0.0224 | -2.14 |
| 000819 | 0.0371 | -33.75 | CL | 0.0582 | 17.92 |
| 000895 | 0.0335 | 2.73 | GS | 0.0336 | -12.06 |
| 000999 | 0.0327 | 21.27 | IBM | 0.0254 | -8.14 |
| 002069 | 0.0435 | -20.93 | KO | 0.0407 | 11.82 |
| 300027 | **0.0133** | -28.61 | MCD | 0.0216 | 27.85 |
| 300029 | 0.1669 | -42.98 | NVDA | 0.0261 | 34.91 |
| 300111 | 0.0320 | -34.09 | WMT | 0.0380 | 16.69 |
| 600084 | 0.0346 | -47.25 | AMZN | **0.0150** | 26.49 |
| 600131 | 0.0344 | -34.58 | CSCO | 0.0212 | 5.27 |
| 600158 | 0.0512 | -34.31 | EA | 0.0327 | 45.24 |
| 600171 | 0.0228 | -34.43 | GOOG | 0.0329 | 21.22 |
| 600199 | 0.0277 | -20.77 | INTC | 0.0290 | -0.60 |
| 600275 | 0.0278 | -59.49 | MSFT | 0.0174 | 14.60 |
| 600300 | 0.0426 | -34.16 | NFLX | 0.0224 | 27.17 |
| 600313 | 0.0335 | -38.67 | TSLA | 0.0320 | 47.84 |
| 600467 | 0.0430 | -27.59 | ULTA | 0.0228 | 18.66 |
| 600519 | 0.0207 | 28.17 | WDC | 0.0247 | 29.19 |

Note: The stock test dataset intervals are all from 2017-01-10 to 2017-06-02.

#### 4.1.2 The stock price prediction errors of different neural networks

By comparing the prediction errors of the two stock dataset, 600275 and AMZN, under the six models, the combination with the smallest prediction error can be selected. In Table 4.2, RNN performs best on the 600275 dataset, and LSTM performs best on the AMZN dataset.

Table 4.2 Comparison of the stock prediction error of different neural networks

| Neural Network | Stock 600275 | | | | Stock AMZN | | | |
|---|---|---|---|---|---|---|---|---|
| | MAE | MSE | RMSE | $R^2$ | MAE | MSE | RMSE | $R^2$ |
| MLP | 0.0329 | 0.0017 | 0.0407 | 0.9604 | 0.0172 | 0.0005 | 0.0217 | 0.9591 |
| RNN | **0.0166** | **0.0006** | **0.0237** | **0.9866** | 0.0118 | 0.0002 | 0.0157 | 0.9785 |
| LSTM | 0.0170 | 0.0005 | 0.0240 | 0.9862 | **0.0115** | **0.0001** | **0.0154** | **0.9794** |
| GRU | 0.0239 | 0.0010 | 0.0310 | 0.9769 | 0.0143 | 0.0003 | 0.0183 | 0.9708 |
| BiRNN | 0.0267 | 0.0012 | 0.0340 | 0.9722 | 0.0167 | 0.0005 | 0.0218 | 0.9588 |
| BiLSTM | 0.0278 | 0.0012 | 0.0353 | 0.9702 | 0.0150 | 0.0004 | 0.0201 | 0.9647 |

## 4.2 Prediction Error Cannot Represent the Ups and Downs of Stock Price

### 4.2.1 The ups and downs of stock prediction error in a shallow network model

In order to analyze the relationship between the prediction error and the ups and downs, the stock samples are divided into two categories: the big difference in prediction error and the small difference in prediction errors, to observe the numerical relationship between them and the actual trend state of stocks. Through the performance of the shallow network MLP on different stock datasets, according to the above two states, specific stock examples are found for analysis. It can be seen that, for different stock datasets, the magnitude of the prediction error value cannot reflect its necessary connection with the rising and falling status of the predicted stock price (Table 4.3).

Table 4.3 The relationship between the size of the prediction error of individual stocks and price's ups and downs in the shallow network model

| State | Group | Stock Symbol | MAE | MSE | RMSE | $R^2$ | Return Rate(%) |
|---|---|---|---|---|---|---|---|
| Large PE Difference | 1 | WMT | 0.0387 | 0.0025 | 0.0498 | 0.9376 | 11.39 |
| | | AMZN | 0.0172 | 0.0005 | 0.0217 | 0.9591 | 210.83 |
| | 2 | 002069 | 0.0480 | 0.0035 | 0.0588 | 0.9269 | -2.33 |
| | | 300027 | 0.0149 | 0.0004 | 0.0195 | 0.9489 | -3.15 |
| Small PE Difference | 1 | CSCO | 0.0265 | 0.0017 | 0.0407 | 0.8854 | 1.60 |
| | | EA | 0.0253 | 0.0016 | 0.0396 | 0.9406 | 34.67 |
| | 2 | 600313.ss | 0.0359 | 0.0031 | 0.0558 | 0.9078 | -2.56 |
| | | 600467.ss | 0.0361 | 0.0023 | 0.0477 | 0.9598 | -1.28 |

Note:
1. The stock test dataset intervals are all from 2017-01-10 to 2017-06-02;
2. Take the return rate of the stock test dataset as the no-treatment-control return rate (%);
3. The return rate refers to the blank control rate of return. It is converted based on the actual price rising and falling values in the stock test dataset interval.

Because of the existence of the absolute value constraint phenomenon, the data that leads to the measurement error cannot reflect the model's actual performance. It can be seen that the users of the shallow neural network MLP cannot make rational decisions based on the evaluation of prediction error (Table 4.3).

### 4.2.2 The ups and downs of stock prediction error in a deep network model

In order to analyze the relationship between the prediction error and the ups and downs, the stock samples are divided into two categories: the big difference in prediction error and the small difference in prediction errors, to observe the numerical relationship between them and the actual trend state of stocks. Through the performance of the shallow network BiLSTM on different stock datasets, according to the above two states, specific stock examples are found for analysis. It can be seen that, for different stock datasets, the magnitude of the prediction error value cannot reflect its necessary connection with the rising and falling status of the predicted stock price (Table 4.4).

Table 4.4 The relationship between the size of the prediction error of individual stocks and price's ups and downs in the deep network model

| State | Group | Stock Symbol | MAE | MSE | RMSE | R2 | Return Rate(%) |
|---|---|---|---|---|---|---|---|
| Large PE Difference | 1 | AAPL | 0.0240 | 0.0010 | 0.0323 | 0.9582 | 30.51 |
| | | 600467 | 0.0430 | 0.0033 | 0.0577 | 0.9412 | -27.59 |
| | 2 | 600171 | 0.0228 | 0.0010 | 0.0323 | 0.9651 | -34.43 |
| | | 600300 | 0.0426 | 0.0032 | 0.0566 | 0.9511 | -34.16 |
| Small PE Difference | 1 | MSFT | 0.0174 | 0.0006 | 0.0242 | 0.9355 | 14.60 |
| | | 600171 | 0.0228 | 0.0010 | 0.0323 | 0.9651 | -34.43 |
| | 2 | MCD | 0.0216 | 0.0010 | 0.0324 | 0.9767 | 27.85 |
| | | 600519 | 0.0207 | 0.0007 | 0.0255 | 0.9590 | 28.17 |

Note:
1. The stock test dataset intervals are all from 2017-01-10 to 2017-06-02;
2. Take the return rate of the stock test dataset as the no-treatment-control return rate (%);
3. The return rate refers to the blank control rate of return. It is converted based on the actual price rising and falling values in the stock test dataset interval.

Because of the existence of absolute value constraints, the data that leads to the measured error cannot truly reflect the model's securities trading function. It can be seen that the users of the deep neural network BiLSTM predicting stock price cannot make rational decisions based on the evaluation metric - the prediction error (Table 4.4).

## 4.3 Differences in Stock Trend that Make the Evaluation Results Ineffective

### 4.3.1 The Harm of Directionless Evaluation in the Securities Trading

In the NNSPP trading evaluation of this study, it is found that two different stocks have the smallest prediction error under the same model. However, within the same test dataset, their returns have a huge difference (Table 4.5). It shows that the prediction error can only reflect the degree of deviation from the forecast, but it cannot reflect the trend direction of the test dataset.

Table 4.5 The stock return direction representation of different stock prediction errors under six models

| Item | Stock | | | | | |
|---|---|---|---|---|---|---|
| | 300027 | | | AAPL | | |
| Stock Trend Movement | 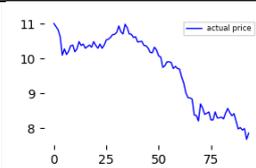 | | | 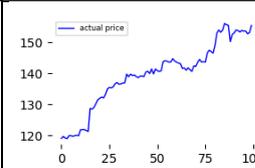 | | |
| Model | MAE | Return Rate | Return Rate Direction | MAE | Return Rate | Return Rate Direction |
| MLP | 0.0149 | | | 0.0203 | | |
| RNN | 0.0105 | | | 0.0140 | | |
| LSTM | **0.0104** | -28.61% | Down | **0.0136** | +30.51% | Up |
| GRU | 0.0117 | | | 0.0152 | | |
| BiRNN | 0.0172 | | | 0.0229 | | |
| BiLSTM | 0.0133 | | | 0.0240 | | |

Note: The test dataset of selected stocks is from 2017-01-10 to 2017-06-02; The prediction error is measured by MAE; The bolded numbers are the minimum prediction errors of the among six models.

It can be seen that the stock 300027 also has the smallest MAE (0.0104) under the model LSTM, but its test dataset's return rate is -28.61%; while the stock AAPL has the smallest MAE under the model LSTM (0.0136), but its test dataset's return rate is +30.51% (Table 4.5). There is a huge difference between the two. The former stock return direction of the test dataset is down, while the latter is up. However, this feature is not reflected by the evaluation indicator MAE. Even NNSPP performance assessors with a little financial knowledge should understand that the stock return direction in this paper is equivalent to the direction of stock rise and fall, which is, the direction of stock price movement.

4.3.2 Statistical Analysis of Model Evaluation Effect using PE

Using MAE, MSE, RMSE, R2 as evaluation metrics, and BiLSTM as the stock price prediction model, after Pearson's correlation analysis of prediction errors and the stock return direction, the Pearson correlation coefficient (PCC) between four types of evaluation metrics, MAE, MSE, RMSE, R2 and the stock return direction in the test dataset interval are -0.2775, -0.2036, -0.2493, 0.1400, respectively. It can be seen that the prediction error has a weak correlation with the stock return direction through model prediction (Fig. 4.1). This is the fundamental reason why the existing PE method cannot truly reflect the overall performance of NNSPP. However, from the experimental results in stocks' actual trading, the stock price trend direction is essential reference information for investors because they need to buy and sell according to their trend direction. The NNSPP that cannot provide predictive results will prevent investors from making correct trading decisions. Such a scenario is similar to why color-blind people who cannot distinguish the red and green colors of traffic lights cannot become pilots! However, it is a pity that many AI developers encountered the problem of NNSPP performance evaluation, but did not realize the seriousness of the problem that the existing PE method cannot represent the stock price direction predicted by NNSPP!

| | MAE | MSE | RMSE | R² | SRD |
|---|---|---|---|---|---|
| **SRD** | -0.2775 | -0.2036 | -0.2493 | 0.1400 | 1 |
| **R²** | -0.4747 | -0.7748 | -0.7240 | 1 | 0.1400 |
| **RMSE** | 0.8753 | 0.9552 | 1 | -0.724 | -0.2493 |
| **MSE** | 0.7031 | 1 | 0.9552 | -0.7748 | -0.2036 |
| **MAE** | 1 | 0.7031 | 0.8753 | -0.4747 | -0.2775 |

Fig.4.1 The heat map of the correlation coefficient matrix between the prediction error and the stock return direction

Note: The stock test dataset of selected stocks is from 2017-01-10 to 2017-06-02; The model used is BiLSTM; The stock return direction is the direction of the stock price in the test dataset; Pearson correlation coefficient is used for correlation analysis, SRD is stock return direction.

From the heat map of the correlation coefficient matrix between the estimated prediction errors for evaluation and the stock return direction of China and the US stocks, it can be seen that there is no correlation between the stock return direction and MAE, MSE, RMSE, and R2 (Fig. 4.1), indicating the quality of the forecast results are not directly related to changes of the stock return direction, so the performance evaluation results using PE cannot be guaranteed to help the practical application of stock trading. In summary, we can conclude that the existing NNSPP performance evaluation methods based on prediction errors cannot truly reflect the model's securities trading performance. This is exactly the financial purpose of why people evaluate NNSPP, and emphasize the importance of predicting the stock price trend direction. Use such evaluation results as the judgment of the NNSPP performance will not only misleads securities investors but also easily brings them substantial financial risks.

5. CONCLUSION

5.1 The Securities Trading Invalidity of NNSPP Evaluation Results is due to the Design Flaws of the Evaluation Method

The importance of the directionality of the financial time series for making trading decisions is self-evident. For example, O. Blaskowitza and H. Herwartz believe that when the prediction model can correctly predict the direction of the financial time series, even if the predicted result is not accurate enough, it can still bring higher investment returns to model users from trading [27]. Khaidem et al. have also pointed out if the direction of stock price changes can be judged under a high probability, and the signal is used as the basis for trading, it can also generate sufficiently high investment returns [42]. From the existing literature, some NNSPP performance evaluators do not understand the stock trading knowledge, and also do not understand the important financial meaning of the stock rising and falling direction, and even do not understand the ability to predict stock price trends is an important part of the performance of NNSPP, which is the main reason that the created NNSPP is out of practical application [4][5][6][7]. In addition, researchers do not set up a blank-control in the experiment, but only compares the prediction error between models. Therefore, the evaluation method of the performance of NNSPP determined by mutual-control-based experiment design will reduce the credibility of the model evaluation results [30].

5.2 The Securities Trading Invalidity of NNSPP Evaluation Results is due to the Absolute Value Constraints of Prediction Errors

The performance of current NNSPP is evaluated based on the PE method, but for specific trading in the securities market, this evaluation method has statistical flaws. As we all know, the PE method takes the absolute value of the prediction error, while the movement of the stock time series is directional. Unfortunately, the prediction error fails to characterize the essential fractal characteristics of the financial time series' inherent rise and fall trends.

The absolute value of any real number is a "non-negative number." $|x| \geq 0$. Since in the predictive system, the estimated target can only be an absolute value and not a negative value, this is the so-called absolute value constraint. Therefore, whether we understand it or not from the angle of the direction or the value reveals an essential property of the absolute value: non-negativity. This simple reason is precisely the source of the problem that the prediction error has no directivity for stocks' ups and downs. As stated above, the stock time series' most crucial property is the directionality of its price movement. At present, the current prediction error used for evaluating the model performance can only reflect the closeness between the model predicted price and the actual market price. However, the core problem that caused the PE method to be questioned was its inability to reflect the direction of the price movement of the stock time series. The existence of this problem caused huge functional defects in the evaluation results of NNSPP. Therefore, lack of interpretability, it is unreliable to judge the performance of the stock price prediction using regression neural networks based only on the high and low of the prediction error value.

ACKNOWLEDGMENT

We thank Cheng Shao from the Stock Exchange of Northeast Securities in Beijing San Li He Branch and Yongchao Jia from the Stock Exchange of Minmetals Securities in Beijing Guang An Men Wai Street Branch for their helpful suggestions in this research. We are also grateful to Lei Zhang from the Stock Exchange of Huaxi Securities in Beijing Zi Zhu Yuan Branch and Sanli Wei from the Stock Exchange of China Merchants Securities in Beijing Che Gong Zhuang West Branch for related information and useful suggestions.

REFERENCES

[1] C. H. Chen, "Neural networks for financial market prediction", Proc. IEEE Int Neural Networks IEEE World Congress Computational Intelligence. Conf., 2: 1199-1202, 1994.
[2] E. Chong, C. Han, and F. C. Park, "Deep learning networks for stock market analysis and prediction: Methodology, data representations, and case studies", Expert Systems with Applications, 83: 187-205, 2017.
[3] H. White, "Economic prediction using neural networks: The case of IBM daily stock returns", Proceedings of the Second IEEE Annual Conference on Neural Networks, 2: 451-458, 1988.
[4] M. Guidolin, S. Hyde, D. McMillan et al., "Non-linear predictability in stock and bond returns: When and where is it exploitable?" International Journal of Forecasting. 25(2): 373-399, 2009.
[5] A. N. Refenes, A. Zapranis, and G. Francis, "Stock performance modeling using neural networks: A comparative study with regression models", Neural Networks, 7(2): 375-388, 1994.
[6] A. B. Suthar, A. Patel and S. M. Parikh, "A comparative study on financial stock market prediction models", The International Journal of Engineering and Science (IJES), 1(2): 188-191, 2012.
[7] I. Kumar, K. Dogra, C. Utreja, et al., "A comparative study of supervised machine learning algorithms for stock market trend prediction", 2018 Second International Conference on Inventive Communication and Computational Technologies (ICICCT), 2018.
[8] S. Selvin, R. Vinayakumar, E. A. Gopalakrishnan, et al., "Stock price prediction using LSTM, RNN and CNN-sliding window model", 2017 International Conference on Advances in Computing, Communications and Informatics (ICACCI), 13-16, 2017.
[9] K. Falcone, "Fractal geometry: mathematical foundations and applications", John Wiley and Sons, 2004.
[10] H. E. Hurst, "The problem of long-term storage in reservoirs", Hydrological Sciences Journal, 1(3): 13-27, 1956.
[11] Y. LeCun, Y. Bengio, and G. Hinton, "Deep learning", Nature, 521(7553): 436-444, 2015.
[12] E. Beede, E. E. Baylor, F. Hersch, et al., "A human-centered evaluation of a deep learning system deployed in clinics for the detection of diabetic retinopathy", In CHI Conference on Human Factors in Computing Systems (CHI'20), 25-30, 2020.
[13] V. Krakovna and F. Doshivelez, "Increasing the interpretability of recurrent neural networks using hidden Markov models", ar Xiv preprint ar Xiv:1606.05320, 2016.
[14] Z. Che, S. Purushotham, R. Khemani, et al., "Distilling knowledge from deep networks with applications to healthcare domain", Annales De Chirurgie, 40(8): 529-532, 2015.
[15] A. Das, H. Agrawal, C. L. Zitnick, et al., "Human attention in visual question answering: Do humans and deep networks look at the same regions?" ar Xiv preprint ar Xiv: 1606.03556, 2016.
[16] D. Martens and F. Provost, "Explaining data-driven document classifications", Mis Quarterly, 38(1): 73-100, 2013.
[17] K. Reing, D. C. Kale, G. V. Steeg, et al., "Toward interpretable topic discovery via anchored correlation explanation", ar Xiv preprint ar Xiv: 1606.07043, 2016.
[18] C. Gallegoortiz and A. L. Martel, "Interpreting extracted rules from ensemble of trees: Application to computer-aided diagnosis of breast MRI", ar Xiv preprint ar Xiv:1606.08288, 2016.
[19] C. Conati, K. Porayska-Pomsta, and M. Mavrikis, "AI in Education needs interpretable machine learning: Lessons from open learner modeling", The Workshop on Human Interpretability in Machine Learning, 2018.
[20] S. M. Kia, "Interpretability of multivariate Brain maps in brain decoding: Definition and quantification", ar Xiv preprint ar Xiv:1603.08704, 2016.
[21] W. Souillard-Mandar, R. Davis, C. Rudin, et al., "Interpretable machine learning models for the digital clock drawing test", ar Xiv preprint ar Xiv: 1606.07163, 2016.
[22] B. Letham, C. Rudin, T. H. Mccormick, et al., "Interpretable classifiers using rules and Bayesian analysis: Building a better stroke prediction model", Annals of Applied Statistics, 9(3): 1350-1371, 2015.
[23] H. Yu, L. R. Varshney, G. E. Garnett, et al., "Learning interpretable musical compositional rules and traces", ar Xiv preprint ar Xiv:1606.05572, 2016.
[24] R. Dash and P. K. Dash, "Efficient stock price prediction using a self evolving recurrent neuro-fuzzy inference system optimized through a modified differential harmony search technique", Expert Systems with Applications 52(15): 75-90, 2016.
[25] Y. Wei and V. Chaudhary, "The directionality function defect of performance evaluation method in regression neural network for stock price prediction", In: DSAA 2020.
[26] C. Stoean, W. Paja, R. Stoean, et al., "Deep architectures for long-term stock price prediction with a heuristic-based strategy for trading simulations", PloS one 14.10: e0223593, 2019.
[27] O. Blaskowitz and H. Herwartz, "On economic evaluation of directional forecasts", International Journal of Forecasting, 27(4): 1058-1065, 2011.
[28] R. N. Elliott, "Nature's law: The secret of the universe", New York, 1946.
[29] W. Bao, J. Yue, and Y. L. Rao, "A deep learning framework for financial time series using stacked autoencoders and long-short term memory", PLoS One, 12(7): e0180944, 2017.
[30] G. Nahler, "Dictionary of pharmaceutical medicine", Springer Vienna, 123-123, 2009.
[31] W. Zhou, X. Y. Zhang, M. F. Xie, et al., "Infrared-assisted extraction of adenosine from radix isatidis using orthogonal experimental design and LC.", Chromatographia, 72(7): 719-724, 2010.
[32] W. G. Cui, X. H. Li, S. B. Zhou, et al., "Investigation on process parameters of electrospinning system through orthogonal experimental design", Journal of Applied Polymer Science, 103(5): 3105-3112, 2007.
[33] R. J. Liang, "Orthogonal test design for optimization of the extraction of polysaccharides from Phascolosoma esulenta and evaluation of its immunity activity", Carbohydrate Polymers, 73(4): 558-563, 2008.
[34] L. K. Wei, X. X. Huang, Z. Z. Huang, et al., "Orthogonal test design for optimization of lipid accumulation and lipid property

in Nannochloropsis oculata for biodiesel production", Bioresource Technology, 147(9): 534-538, 2013.

[35] S. Bermejo and J. Cabestany, "Oriented principal component analysis for large margin classifiers", Neural Networks, 14(10): 1447-1461, 2001.

[36] R. J. Hyndman and A. B. Koehler. "Another look at measures of forecast accuracy", International Journal of Forecasting, 22(4): 679-688, 2006.

[37] C. J. Willmott and K. Matsuura, "Advantages of the mean absolute error (MAE) over the root mean square error (RMSE) in assessing average model performance", Climate Research, 30(1): 79-82, 2005.

[38] R. J. Hyndman and G. Athanasopoulos, "Forecasting: principles and practice", OTexts, 2018.

[39] N. R. Draper and H. Smith, "Applied regression analysis", John Wiley & Sons, 1998.

[40] K. Pearson, "Notes on regression and inheritance in the case of two parents", Proceedings of the Royal Society of London, 58: 240-242, 1895.

[41] D. Hinkle, W. Wiersma, and S. G. Jurs, "Applied statistics for the behavioral sciences", Houghton Mifflin College Division, 2003.

[42] L. Khaidem, S. Saha, and S. R. Dey, "Predicting the direction of stock market prices using random forest", Applied Mathematical Finance, 2016.